\def\refitem{\par\parskip 0pt\noindent\hangindent 20pt}
\def\spose#1{\hbox to 0pt{#1\hss}}
\def\approxgt{\mathrel{\spose{\lower 3pt\hbox{$\sim$}}
	\raise 2.0pt\hbox{$>$}}}
\def\cm{{\rm\thinspace cm}}
\def\erg{{\rm\thinspace erg}}
\def\s{{\rm\thinspace s}} 
\def\Msun{\hbox{$\rm\thinspace M_{\odot}$}}
\def\ergpcmsqps{\hbox{$\erg\cm^{-2}\s^{-1}\,$}}
\def\ergps{\hbox{$\erg\s^{-1}\,$}}

\documentstyle[psfig]{mn}
\begin{document}
\hsize=6truein

\title{Variability of the extreme z=4.72 blazar, GB~1428+4217}

\author[]
{\parbox[]{6.in} {A.C.~Fabian$^1$, A. Celotti$^{1,2}$, G. Pooley$^3$,
K.  Iwasawa$^1$, W.N.~Brandt$^4$, R.G. McMahon$^1$ and M.D. Hoenig$^1$\\
\footnotesize
1. Institute of Astronomy, Madingley Road, Cambridge CB3 0HA \\
2. S.I.S.S.A., via Beirut 2-4, 34014 Trieste, Italy\\ 
3. Cavendish Laboratories, Madingley Road, Cambridge CB3 0HE  \\ 
4. Department of Astronomy and Astrophysics, The Pennsylvania State
University, 525 Davey Lab., University Park, PA 16802, USA\\
\\}}
\maketitle

\begin{abstract}
We report X--ray and radio variability of GB 1428+4217 which confirm
its blazar nature. IR observations reveal a powerful optical--UV
component, not obscured by dust, which is suggestive of the presence
of a billion solar mass black hole, already formed by $z\sim 5$. A
detailed comparison of the broad band spectral properties of GB
1428+4217 with those of nearby blazars shows it to be extreme, but
nevertheless consistent with the trend found for nearby sources.
\end{abstract}


\section{Introduction}
At redshift $z=4.72$, GB 1428+4217 is the most distant X-ray source
known (Hook \& McMahon 1997; Fabian et al 1997; 1998). Its bright
X-ray flux, of about $3\times 10^{-12}\ergpcmsqps$ in the 2--10~keV
band, and radio flux, of about 100~mJy at 15~GHz, strongly suggest
that the object is a blazar pointed toward us. The previous
observations did not however clearly show the flux variability which
is common to blazars. We have therefore observed GB~1428+4217 again in
X--rays with the ROSAT HRI and have monitored it for several months
with the Ryle radio telescope. We report here on our discovery of the
expected X--ray and radio variability. We also now make a detailed
comparison of its properties with those of nearby blazars.

\section{ROSAT and Radio variability data}

GB~1428+4217 was observed 4 times with the ROSAT HRI during 1997
December and 1998 January. The source is very clearly detected each
time and shows significant variability (see Fig.~1 where we include
the earlier HRI point from 1996). It varied by a factor of about two
over a timescale of two weeks (or less), which corresponds to less
than 2.5 days in the restframe of the source. We note that previous
ASCA and serendipitous ROSAT detections of the source give a flux
consistent with that of 1996 July.

\begin{figure}
\centerline{\psfig{figure=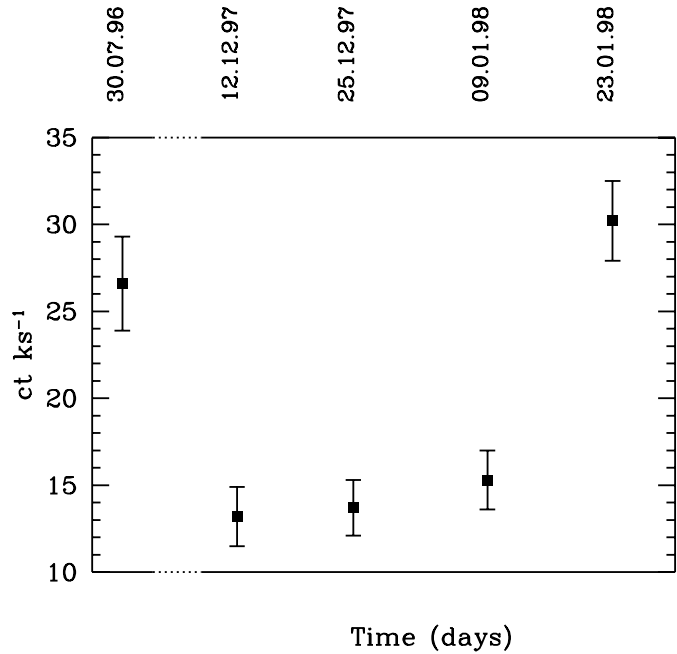,width=0.5\textwidth,angle=0}}
\caption{Significant count rate variability can be inferred between the
five different ROSAT HRI observations during 1996, 1997 and
1998. Dates are indicated on the top of the plot, and the
corresponding exposure times for the 1997 and 1998 datasets are 4.8,
5.9, 5.8 and 6.2 ks, respectively.  As a reference point, 27 ct
ks$^{-1}$ correspond to 10$^{-12}$ erg cm$^{-2}$ s$^{-1}$ in the
(0.1--2.4 keV) ROSAT band.}
\end{figure}

\begin{figure}
\centerline{\psfig{figure=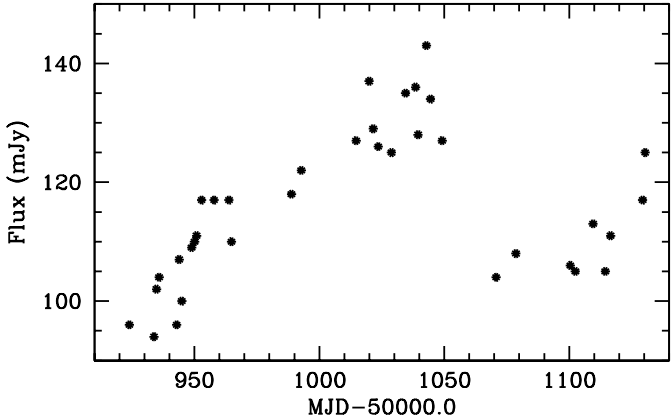,width=0.5\textwidth,angle=0}}
\caption{Radio light curve at 15 GHz, showing the significant 
variability of GB 1428+422.}
\end{figure}

The Ryle Telescope at Cambridge was used to monitor the flux density
of GB~1428+422 at 15 GHz on about 50 occasions between 1998 April and
December (unfortunately much later than the HRI pointing). These
observations, usually of short duration (typically $<$ 1 h), were made
during gaps in the regular schedule of the telescope. A more detailed
description of the observing technique is given in Pooley \& Fender
(1997).  The flux--density scale of the observations was established
by a nearby observation of either 3C~48 or 3C~286. Similar datasets
have shown that the overall calibration in such cases has an
r.m.s. scatter of less than 3 per cent.

In Fig.~2 the flux density of GB~1428+422 is plotted: significant
variability can be detected with an amplitude of $\sim$ 40 per cent
over about three months, and $\sim$ 15 per cent on timescale of ten
days.

We therefore conclude that the significant variability on short
timescales detected both in the X--ray and radio bands strongly
confirms the identification of GB~1428+4217 with a blazar. In the
following we will then consider the variability and spectral
properties of this source, and compare them with those of nearby
blazars, with the aim to gain insight both on the physics and
evolutionary behavior of this class of AGN.

\section{Infrared Observations}

In addition to the broad band B and R magnitudes (and monochromatic
continuum flux at a rest frame wavelength of 1500\AA) reported by Hook
\& McMahon (1998), we present here $J$, $H$ and $K$ observations. They
were carried out on 1997 March 26 with the United Kingdom Infra-Red
Telescope (UKIRT) using the $256^{2}$ InSb array based camera IRCAM3
in the 0.28arcsec/pixel mode with exposure times per waveband of 900
s.  The data underwent dark subtraction, flat-fielding, and mosaicing
of the dithered images using the Starlink IRCAMDR software.
Photometry was carried out using apertures of diameter 5 arcsec
calibrated against similarly analyzed standard stars from Casali \&
Hawarden (1992). In Table~1, we list the observed magnitudes and
derived fluxes.

The spectral index over the range covered by the optical and IR data
is $\sim$0.0 ($F_{\nu} \propto \nu^{-\alpha}$). This is bluer than the
canonical $\alpha_{\rm uv}$ of 0.7 (eg. Fall, Pei \& McMahon 1989)
which means that there is no evidence for reddening over the rest
frame spectral range 1500--4000\AA.

\begin{table*}
\begin{center}
\label{obs}
\caption{Infrared observations}
\begin{tabular}{cccccc}
\\
\hline\hline
\\
  \multicolumn{1}{c}{Band} 
& \multicolumn{1}{c}{Wavelength}
& \multicolumn{1}{c}{Wavelength}
& \multicolumn{1}{c}{Magnitude}
& \multicolumn{1}{c}{Zeropoint$^\dagger$}
& \multicolumn{1}{c}{Flux}
\\
&Observed ($\mu m$)& Rest (\AA)& 
&$\rm erg~cm^{-2}~Hz^{-1}$  &$\rm mJy
$  \\
\\
\hline
\\
J&1.22 &2140\AA &18.6$\pm$0.1&$1.57\times10^{-20}$ &$5.7\times10^{-2}$ \\ 
H&1.63 &2860\AA &18.0$\pm$0.1&$1.02\times10^{-20}$ &$6.4\times10^{-2}$ \\
K&2.19 &3840\AA &17.5$\pm$0.1&$6.36\times10^{-21}$ &$6.3\times10^{-2}$ \\
\\
\hline
\end{tabular}
\end{center}
\noindent Notes: $\dagger$ Effective wavelengths and zero point fluxes
on the $\alpha$-Lyrae magnitude system from Bessel \& Brett (1988).
\end{table*}

\section{Variability constraints}

The spectral energy distribution (SED) of this source already pointed
towards its identification with a blazar (Fabian et al. 1997, 1998).
In particular, Fabian et al. (1998) showed that the (poorly sampled
and not simultaneous) SED of GB~1428+4217 can be accounted for as
non--thermal synchrotron and inverse Compton emission from a
relativistically moving source, forming two broad peaks in $\nu
F(\nu)$, as characteristic of blazars.  However, no constraints on its
size were available at the time. The X--ray variability timescale
inferred from our ROSAT observations sets instead a significant upper
limit on the dimension. We therefore re-consider the modeling of the
SED and find that a broad band energy distribution and flat X--ray
spectrum consistent with the data can still be found adopting a simple
homogeneous model. As an example of the results obtained, in Fig.~3 we
show the SED from one of the specific models (see caption) proposed by
Fabian et al. (1998), where the intrinsic dimensions and Doppler
factor are of order $R\sim 5\times 10^{16}$ cm and $\delta \sim 20$,
respectively.  As already pointed out by Fabian et al.  the parameters
inferred from the modeling are globally consistent with those deduced
for larger samples of blazars at lower redshifts (Ghisellini et
al. 1998).

A further hint that beaming is involved is given by the rate of change
of luminosity, $\Delta L/\Delta t \approxgt 5\times 10^{41}$ erg
s$^{-2}$.  The simple efficiency limit (Fabian 1979; Brandt et al.
1999) then yields a radiative efficiency for the source $\approxgt$ 20
per cent if only the luminosity in the ROSAT band is considered, and
about ten times higher if one consider the total X--ray luminosity.

\begin{figure}
\centerline{\psfig{figure=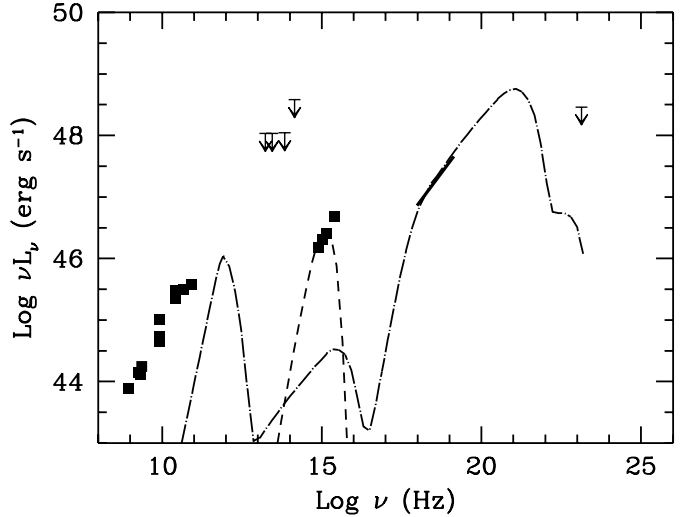,width=0.5\textwidth,angle=0}}
\caption{Broad band spectral energy distribution (data are not 
simultaneous).  Superposed is the predicted SED from a homogeneous
model, for a Doppler factor $\delta\sim 20$ and an intrinsic source
typical size $R\sim 5\times 10^{16}$ cm, while the other model
parameters are the same as in Fabian et al. (1998) (EC model).  The
dashed peaked component represents the possible contribution by an
external soft radiation field, which provides the seed photons for the
inverse Compton high energy emission. The X--ray spectrum is derived
from the ASCA data (flux level comparable with the ROSAT HRI 1996
observation).}
\end{figure}

\section{Broad band spectral energy distribution}

A closer comparison can now be made between the spectral properties of
GB~1428+4217 and those of nearby blazars. Differences can give
important hints on the evolution in the intrinsic or environmental
properties of radio--loud AGN.

Here we consider as quantitative SED indicators the broad band
spectral indices\footnote{where the quasar restframe radio, optical
and X--ray monochromatic fluxes are calculated at 5 GHz, 5500 \AA\ and
1 keV.}, $\alpha_{\it ro}$, $\alpha_{\it ox}$ and $\alpha_{\it rx}$.

Because of the uncertainties due to the flux variability we estimate
the range spanned by $\alpha_{\it ro}$, $\alpha_{\it ox}$ and
$\alpha_{\it rx}$, by considering the extremes of the observed radio
and X--ray flux ranges. The two areas in Fig.~4 are representative of
the intervals obtained.  In this same figure, the spectral indices
derived for GB~1428+4217 are compared with those obtained for complete
samples of BL Lac objects and flat spectrum radio quasars (FSRQ; see
Fossati et al. 1998 for details).  Other two $z>4$ blazars,
GB1508+5714 at $z=4.3$ (Moran \& Helfand 1997) and RXJ 1028.6-0844 at
$z=4.28$ (Zickgraf et al. 1997), appear similar to GB~1428+4217,
subject to the uncertainty in the necessary optical K-correction for
these objects (we adopt an optical spectral index for them of 0.7).

These three sources lie apart from the nearby FSRQ reported.  They do
however follow the general trends which hold for the entire blazar
class.  It has been pointed out that if blazars are considered
according to their total (and radio) power, then the
presence/luminosity in (broad) emission lines correlates with the
shape of the SED. In particular, the position in energy of the two
broad continuum peaks and their relative intensity (power in the low
energy component with respect to that in the high energy one) decrease
with increasing total source power. This also translates into an
increase of the $\alpha_{\rm ro}$ and $\alpha_{\rm rx}$ spectral
indices, and a flattening of the latter at the highest luminosities
(see Fossati et al. 1998).  Note indeed that GB~1428+4217 follows
these trends in the spectral index plane as predicted by its radio
power.  An interesting consequence is that one would expect the
synchrotron peak to be located in the mm band, well below the apparent
optical peak of the (sparse) SED shown in Fig.~3.

\begin{figure}
\centerline{\psfig{figure=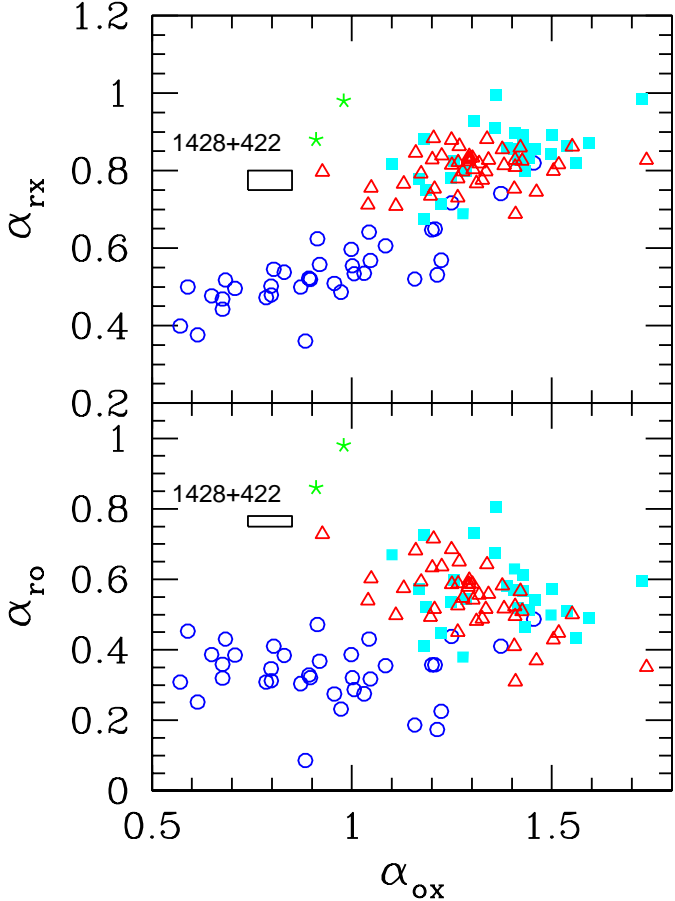,width=0.5\textwidth,angle=0}}
\caption{Broad band spectral indices for complete samples of radio--loud 
quasars (squares) and BL Lac objects (open circles and triangles),
from the Wall \& Peacock 2 Jy sample (Wall \& Peacock 1985; Padovani
\& Urry 1992), the 1 Jy and Einstein Slew survey BL Lac samples 
(Stickel et al. 1991; Stickel, Meisenheimer \& K\"uhr 1994; Perlman et
al. 1996), respectively.  GB~1428+4217 spectral indices, estimated
considering the variable radio and X--ray observed fluxes, span the
range indicated by the boxes. The stars refer to the other two $z>4$
objects (GB~1508$+$5714 and RXJ~1028.6-0844), whose optical spectral
index is assumed $\sim 0.7$. }
\end{figure}

Although the behaviour of GB~1428+4217, and the other two $z>4$
blazars, fit the blazar scenario just described, the X--ray
luminosities of the sources still exceed those predicted by the above
correlations; the values of $\alpha_{\rm ox}$ are significantly
flatter than expected for the given source powers. Two caveats should
however be remembered, namely the critical role of variability when
discussing properties of blazars, and the possible influence of
selection effects in detecting X--ray emission from GB~1428+4217.
Indeed, a similar object with less extreme X--ray luminosity (fully
consistent with the trends discussed above) would not have been
followed up so readily in the X--ray band.  Similar considerations
apply for the other two high redshift objects, although we stress that
the uncertainties in the (radio, optical and X--ray) spectral slopes
of these two sources is large.

We now consider the implication of these results. Within the most
widely accepted blazar scenario the two peaks of the SED are
interpreted as due to synchrotron (the low energy component) and
inverse Compton (the high energy one) processes. Although the nature
of the seed photons [internal/synchrotron (SSC) vs external radiation
field (EC)] is still not fully settled, evidence has been found that
the source power is directly linked to the magnitude of the external
radiation field, thus implying the increasing importance of the EC
mechanism over the SSC one with increasing power. This also means more
effective radiative cooling of electrons (by inverse Compton) - and
thus a possible interpretation of the lowering of the energy of
particles which emit at the peaks - and an increase in the relative
importance of the Compton spectral component (Ghisellini et al. 1998).

If so, GB~1428+4217 (and maybe the other two high redshift quasars)
represents a powerful source with an intense (or even extreme)
external radiation field. This does not seem to be supported by the
lack of very luminous emission lines (the photon field of which could
be an important contributor to the external seed field).

However, there is interesting evidence for an intense optical--UV
continuum flux, which cannot be easily interpreted as non--thermal
emission. It seems plausible that this component (and that responsible
for the intense Compton emission) have a nuclear origin.  In fact no
other radiation field (e.g. any plausible star cluster, or the cosmic
microwave background radiation) seems to be energetically relevant. It
is possible that this component is analogous to the excess optical--UV
emission in nearby quasar and might be ascribed to thermal dissipation
from accreting material, which would itself contribute to the local
external radiation field.

Independently of its origin it should be stressed that the detection
of this optical component sets extremely tight limits to the presence
of any dust along the line of sight to this high $z$ source.

\section{Discussion and conclusions}

Evidence for X--ray and radio variability in GB~1428+4217 confirms the
blazar nature of this quasar.

Interestingly and perhaps surprisingly, the properties of such an
extreme high redshift source seem to fit globally the scenario for low
redshift blazars. Although no conclusive evidence of peculiar
intrinsic or environmental conditions can be found, there is some
indication of even more extreme Compton cooling which might be
associated for example with an unusually high external radiation
field.

While no conclusions can be drawn on this basis, the issue remains
open. Are all high redshift blazars characterized by such high
(X--ray) luminosities?  In Fig.~5 we show the broad band energy
distributions of GB~1428+4217 and the two other $z>4$ radio--loud
blazars, GB~1508+5714 and RXJ 1028.6--0844. These are also
characterized by a very flat X--ray spectral index and the broad band
properties of luminous objects, and at least GB~1508+5714 might share
with GB~1428+4217 such an extreme X--ray brightness. Broad band
spectral coverage and variability studies of a significant number
(possibly a complete sample) of such `primordial' blazars are
required.

We finally note that if indeed the detected optical--UV flux can be
ascribed to a thermal component produced in the accreting process, a
bolometric luminosity of $\sim$ 10$^{47}\ergps$ requires the presence
of a black hole with billion solar masses, if accreting at the
Eddington rate, which has to be formed by $z\sim 5$ (Efstathiou \&
Rees 1988).  Furthermore, one can speculate that if the black hole
mass is related to the mass of the galactic bulge according to the
relation suggested by Maggorian et al. (1998), a $\sim$ 10$^{11}\Msun$
bulge component has also to be present. Clearly the observational
confirmation or rejection of these possibilities is of crucial
importance for the study of galaxy and black hole formation and their
mutual relationship.

\begin{figure}
\centerline{\psfig{figure=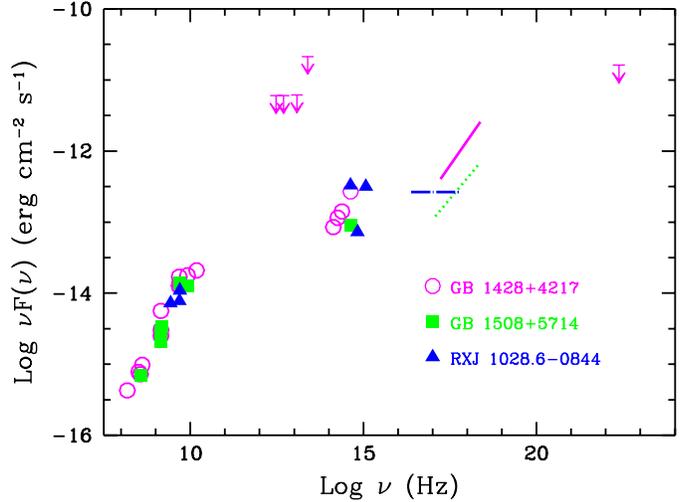,width=0.5\textwidth,angle=0}}
\caption{The SED of three $z\approxgt 4$ radio--loud quasars are compared. 
The objects show a similar broad band behavior, with extreme flat
X--ray spectral indices, low energy synchrotron and (plausibly)
inverse Compton peaks. The continuous, dotted and dot--dashed lines
reproducing the X--ray spectra refer to GB~1428+4217, GB~1508+5714 and
RXJ~1028.6-0844, respectively. The X--ray spectrum of GB~1428+4217 is
from the ASCA data.}
\end{figure}

\section*{Acknowledgments}

We thank Gabriele Ghisellini for the use of the code for the
homogeneous emission model.  The Royal Society (ACF, RGM), the Italian
MURST (AC), PPARC (KI) and the NASA LTSA Program (WNB) are thanked for
financial support. This research was supported in part by the National
Science Foundation under Grant No. PHY94-07194 (AC).

\section*{References}

\refitem Bessell M.S., Brett J.M., 1988, PASP, 100, 1134
\refitem Brandt W.N., Boller Th., Fabian A.C., Ruszkowski M., 1999, 
 MNRAS, submitted
\refitem Casali M.M., Hawarden T.G, 1992, JCMT-UKIRT Newsletter, 3, 33
\refitem Efstathiou G., Rees M.J., 1988, MNRAS, 230, 5P
\refitem Fabian A.C., 1979, Proc. Roy. Soc. A, 336, 449 
\refitem Fabian A.C., Brandt W.N., McMahon R.G., Hook I., 1997,
 MNRAS, 291, L5
\refitem Fabian A.C., Iwasawa K., Celotti A., Brandt W.N., McMahon R.G., 
 Hook I., 1998, MNRAS, 295, L25
\refitem Fall S.M., Pei Y.C., McMahon R.G., 1989, ApJ, 341, L5
\refitem Fossati G., Maraschi L., Celotti A., Comastri A., Ghisellini G., 
 1998, MNRAS, 299, 433
\refitem Ghisellini G., Celotti A., Fossati G., Maraschi L., Comastri 
 A., 1998, MNRAS, 301, 451
\refitem Hook I.M., McMahon R.G., et al, 1994, MNRAS, 273, L63
\refitem Hook I.M., McMahon R.G., 1997, MNRAS, submitted 
\refitem Magorrian J., et al., 1998, AJ, 115, 2285
\refitem Moran E.C., Helfand D.J., 1997, ApJ, 484, L95 
\refitem Padovani P., Urry C.M., 1992, ApJ, 387, 449
\refitem Perlman E.S., et al., 1996, ApJS, 104, 251
\refitem Pooley G.G., Fender R.P., 1997, MNRAS, 292, 925
\refitem Stickel M., Fried J.W., K\"uhr H., Padovani P., Urry C.M., 1991, 
 ApJ, 374, 431
\refitem Stickel M., Meisenheimer K., K\"uhr H., 1994,
 A\&AS, 105, 211
\refitem Wall J.V., Peacock J.A., 1985, MNRAS, 216, 173
\refitem Zickgraf F.-J., Voges W., Krautter J., Thiering I., Appenzeller I., Mujica R., Serrano A., 1997, A\&A, 323, L21

\end{document}